\documentclass[acmsmall]{acmart} 
\setcopyright{rightsretained}
\acmDOI{10.1145/3643746}
\acmYear{2024}
\copyrightyear{2024}
\acmSubmissionID{fse24main-p338-p}
\acmJournal{PACMSE}
\acmVolume{1}
\acmNumber{FSE}
\acmArticle{20}
\acmMonth{7}
\received{2023-09-28}
\received[accepted]{2024-01-23}

\AtBeginDocument{%
  }

\usepackage{tikz}
\usepackage{amsmath}

\usepackage{listings}
\usepackage{array}
\usepackage{subcaption}
\usepackage{wrapfig}
\newcolumntype{L}[1]{>{\raggedright\let\newline\\\arraybackslash\hspace{0pt}}m{#1}}
\newcolumntype{C}[1]{>{\centering\let\newline\\\arraybackslash\hspace{0pt}}m{#1}}
\newcolumntype{R}[1]{>{\raggedleft\let\newline\\\arraybackslash\hspace{0pt}}m{#1}}
\usepackage{algorithm}
\usepackage{graphicx}
\usepackage{textcomp}
\usepackage{xcolor}
\usepackage{multirow}
\usepackage{xurl}
\usepackage{colortbl}

\usepackage[toc,page]{appendix}
\usepackage{syntax}
\usepackage{semantic}
\usepackage{proof}
\usepackage{amsmath}
\usepackage{mathtools}
\usepackage{cleveref}
\crefformat{footnote}{#2\footnotemark[#1]#3}
\usepackage{footmisc}

\usepackage{algpseudocode}
\algrenewcommand{\algorithmiccomment}[1]{{\small\hfill$\triangleright$ #1}}
\algrenewcommand\algorithmicindent{1em}

\algnewcommand\algorithmicswitch{\textbf{switch}}
\algnewcommand\algorithmiccase{\textbf{case}}
\algnewcommand\algorithmicassert{\texttt{assert}}
\algnewcommand\Assert[1]{\State \algorithmicassert(#1)}%
\algdef{SE}[SWITCH]{Switch}{EndSwitch}[1]{\algorithmicswitch\ #1\
\algorithmicdo}{\algorithmicend\ \algorithmicswitch}%
\algdef{SE}[CASE]{Case}{EndCase}[1]{\algorithmiccase\ #1}{\algorithmicend\ \algorithmiccase}%

\algdef{SE}[DOWHILE]{DoWhile}{EndDoWhile}{\algorithmicdo}[1]{\algorithmicwhile\ #1}%

\definecolor{gray}{gray}{0.3}
\definecolor{light-gray}{gray}{0.9}

\newcommand{\toolname}{\textsf{krepair}}

\begin{document}

\title{Maximizing Patch Coverage for Testing of Highly-Configurable Software without Exploding Build Times}


\author{Necip Fazıl Yıldıran}
\orcid{0000-0001-7297-5462}
\affiliation{%
  \institution{University of Central Florida}
  \city{Orlando}
  \country{USA}
}
\email{yildiran@knights.ucf.edu}

\author{Jeho Oh}
\orcid{0000-0002-5599-268X}
\affiliation{%
  \institution{University of Texas}
  \city{Austin}
  \country{USA}
}
\email{jeho.oh@utexas.edu}

\author{Julia Lawall}
\orcid{0000-0002-1684-1264}
\affiliation{%
  \institution{Inria}
  \city{Paris}
  \country{France}
}
\email{julia.lawall@inria.fr}

\author{Paul Gazzillo}
\orcid{0000-0003-1425-8873}
\affiliation{%
  \institution{University of Central Florida}
  \city{Orlando}
  \country{USA}
}
\email{paul.gazzillo@ucf.edu}


\begin{abstract}
The Linux kernel is highly-configurable, with a build system that takes a configuration file as input and automatically tailors the source code accordingly.
Configurability, however, complicates testing, because different configuration options lead to the inclusion of different code fragments.
With thousands of patches received per month, Linux kernel maintainers employ extensive automated continuous integration testing.
To attempt patch coverage, i.e., taking all changed lines into account, current approaches either use configuration files that maximize total statement coverage or use multiple randomly-generated configuration files, both of which incur high build times without guaranteeing patch coverage.
To achieve patch coverage without exploding build times, we propose \toolname{}, which automatically repairs configuration files that are fast-building but have poor patch coverage to achieve high patch coverage with little effect on build times.
krepair works by discovering a small set of changes to a configuration file that will ensure patch coverage, preserving most of the original configuration file's settings.
Our evaluation shows that, when applied to configuration files with poor patch coverage on a statistically-significant sample of recent Linux kernel patches, krepair achieves nearly complete patch coverage, 98.5\% on average, while changing less than 1.53\% of the original default configuration file in 99\% of patches, which keeps build times 10.5x faster than maximal configuration files.

\end{abstract}

\ccsdesc[500]{Theory of computation~Program analysis}
\ccsdesc[500]{Software and its engineering~Software maintenance tools}
\ccsdesc[300]{Software and its engineering~Software testing and debugging}

\keywords{software configuration, build systems, static analysis}

\maketitle

\section{Introduction}

The Linux kernel is a prototypical example of a highly-configurable system.  Users can adapt the Linux kernel to virtually endless combinations of hardware and software requirements by simply selecting configuration options, with no additional programming~\cite{fosdbook,gazzillo20,ieeeconfig}. This high degree of configurability allows the Linux kernel to be used in very diverse environments, including all of the top 500 supercomputers \cite{top500}, 40\% of servers~\cite{usingunix}, and the majority of Internet-of-Things devices~\cite{iotsurvey18}.  Nevertheless, this degree of configurability complicates testing, because different configuration options lead to the inclusion of different code fragments and thus different runtime behaviors.  Configurability is especially challenging when the software is rapidly changing, as changes must be validated with respect to software configurations that actually do include the changed code.  The Linux kernel receives thousands of patches per month, and automated continuous integration testing is extensively used to cope with this rate of change.  To try to achieve {\em patch coverage}, i.e., that all changed lines are taken into account, current continuous integration testing approaches either use configuration files that select as many configuration options as possible (for the Linux kernel, \texttt{make allyesconfig}) or use multiple randomly generated configuration files (\texttt{make randconfig}), both of which lead to high build times without guaranteeing success.

State-of-the-art approaches to generating configuration files target increasing feature-interaction coverage or statement coverage, but are not designed for patch coverage.
Approaches targeting feature-interaction coverage systematically test many combinations of features~\cite{taitest2002,cohen96}, e.g., all pairs of features or all triples of features.  But such approaches do not scale to testing software with large numbers of configuration options~\cite{10sampling}, and even for smaller systems they generate thousands of configuration files, requiring enormous resources to build and test continuously.
And feature-interaction coverage does not guarantee patch coverage~\cite{10sampling}.

Statement-coverage approaches, in contrast, seek to cover the most code with the fewest configuration files~\cite{10sampling}, which results in high build times, and can still fail to cover patches.
Indeed, the configuration file obtained using \texttt{allyesconfig} takes more than ten times longer to build than the Linux kernel's default configuration (obtained using \texttt{make defconfig}), and the very large size of \texttt{allyesconfig} means that it is not suitable for booting on some machines~\cite{oracle}.
Statement-coverage approaches are thus highly resource intensive for continuous integration testing, which needs to test hundreds of patches several times a day.
Moreover, when examining build-test reports from the Linux kernel 0-day build testing service~\cite{intel0day}, the large majority (63\%) are randomly-chosen configurations (\texttt{randconfig}) compared to many fewer reports of \texttt{allyesconfig} (15\%) (Section~\ref{sec:overview}).

To achieve patch coverage while preserving the original configuration file and its build times, we propose to construct configurations that are targeted to the specific changes found in a given patch.
We introduce a new algorithm, called \toolname{}, to solve the problem of generating configuration files for efficient continuous integrating testing of highly-configurable software.
It works by \mbox{\emph{repairing}} a user-provided configuration file to ensure patch coverage without resorting to maximal configurations, and preserves most of the original configuration file's settings.
For instance, Linux's small default configuration (\texttt{make defconfig}) rarely covers the code in patches but builds relatively quickly.
After repairing by \toolname{} for a given patch, the repaired default configuration almost always covers the patch with only marginal additional build time, while in most cases, \toolname{} only takes a few minutes to find a covering configuration file.
This approach thus retrofits existing continuous integration testing for highly-configurable software to provide high patch coverage with little additional cost, since it repairs any existing configuration files already used or generated by testers.

\toolname{} works by discovering, using automated reasoning, a small set of changes to a configuration file that will ensure patch coverage.
It first collects a set of patch coverage constraints for all changed lines of code.
This step draws on statement-coverage approaches~\cite{tartler2014}, using existing line coverage constraint extractors~\cite{oh21,fse17,gazzillo-pldi12} and building on the VAMPYR algorithm~\cite{tartler2014} to find a set of patch-covering constraints for a given patch.
The challenge for \toolname{} is to combine these constraints with a test platform's existing configuration file, which often has low patch coverage but fast build times, without introducing contradictions; such contradictions indeed often arise because of the many complex dependencies among the Linux kernel configuration options.
To overcome this challenge, \toolname{} uses an automated theorem prover to detect which configuration file settings cause contradictions with the patch coverage constraints and removes these settings, little by little, until the resulting configuration file satisfies the constraints.
Then, it repopulates missing configuration settings by querying the prover for a solution that preserves the patch constraints.
Since automated theorem proving is expensive, \toolname{} employs several optimizations to reduce the number of calls to the prover.
When there are mutually-exclusive changes in a patch, i.e., no one configuration file can cover the patch, \toolname{} detects this and generates a small set of repaired configuration files that collectively cover the whole patch.  In practice, 97\% of patches we have tested produced just one configuration file.
We have implemented \toolname{} in Python as a command-line tool that works on the Linux build system, a build system that is also used by other low-level systems software, such as BusyBox~\cite{busybox} and coreboot~\cite{coreboot}. 

We evaluate \toolname{} by measuring how well it ensures that configuration files cover patches while keeping the build times fast enough for continuous integration testing.
We use a statistically significant sample of patches from one full of year of about 71,000 patches resulting in a sample of 507 patches.
We quantify patch coverage as the number of removed or added lines included by the build configuration divided by the total number of removed or added lines in the patchfile.
To measure patch coverage, we intercept the build system to check whether the files and lines of the patch have been included in the build, then we compare the coverage of each patch before and after repair.
To measure build time, we build the entire kernel from scratch on an AMD EPYC compute server using the configuration file and record the wall clock time.
We compare the patch coverage and build time against the Linux default configuration and its statement maximizing configuration file \texttt{allyesconfig}.
The set of configuration files generated by \toolname{} achieves 98.5\% patch coverage on average, compared to 21.7\% for the default configuration file, \texttt{defconfig}.  \toolname{}'s configuration files even have higher coverage than \texttt{allyesconfig} on average, which covers 88.5\%.  But \toolname{}'s set of \texttt{defconfig}-based configuration files are 10.5x faster to build than \texttt{allyesconfig} and comparable in build time to \texttt{defconfig}.  In short, \toolname{} achieves the patch coverage of statement-covering approaches without the cost in resources, taking only a small fraction of the build time, while \toolname{} itself finds a patch-covering configuration file in a few minutes in most cases.

\toolname{} achieves fast build times, because it preserves most settings from its input configuration file while still covering patches.
We show that in 99\% of patches, it only changes 1.5\% or fewer configuration options to achieve patch coverage.
Additionally, since random configuration testing is used by some of the largest industrial continuous testing infrastructures~\cite{intel0day,fixstats}, we also measure how much patch coverage such testing can achieve.
We show that a single random configuration file only achieves 29.2\% patch coverage on average, while adding more random configuration files has diminishing coverage returns, plateauing at around 75\% with 10 random configuration files.  Moreover, using multiple random configuration files to achieve patch coverage increases build time, since each random configuration file needs to be built individually for testing.
We even find some build errors that were overlooked when the patches were integrated into the Linux kernel, showing that \toolname{} complements existing testing approaches.
We describe the 18 build errors found by repaired configuration files, including 2 errors that had not yet been fixed, one of which has already had our patch accepted by the Linux developers.

This paper makes the following contributions:
\begin{itemize}
\item An algorithm that automatically repairs existing configuration files to cover patches with little effect on build times (Section~\ref{sec:algorithms}).
\item The implementation of \toolname{}, with caching to improve performance (Section~\ref{sec:implementation}).
\item An evaluation of \toolname{} for patch coverage, build times, and configuration preservation, with a comparison to state\-ment-coverage maximizing and random configuration-file generation approaches (Section~\ref{sec:evaluation}).
\end{itemize}
\vspace{3em}

\section{Background}
\label{sec:overview}

When testing tools do not ensure patch coverage, they cannot exhaustively test changes to the code.  For instance, \texttt{syzbot} performs continuous testing of the Linux kernel using the \texttt{syz\-kaller}~\cite{syzkaller} fuzz-tester and was responsible for the majority of credited reports to the release v4.9~\cite{fixstats55}.
But it relies on a small, fixed set of configuration files with the configuration options necessary to run syzbot~\cite{syzkaller}.
These configurations provide no assurance that code in new patches gets compiled before testing.
A memory leak in the Linux kernel~\cite{syzbotreport} that \texttt{syzkaller} can detect~\cite{syzbotusbfuzzsupport} remained in the kernel for months, because the configuration option controlling inclusion of the buggy code was not enabled.  \texttt{syzkaller} only found the bug months later, after the configuration option happened to be included by the default configuration in a later version of Linux~\cite{syzbotenabledconfig}.

To understand the configurations that the Linux kernel developer community
considers to be useful to test, we study the e-mail history\footnote{\url{https://web.archive.org/web/20221023104058/https://lists.01.org/hyperkitty/}} of the Linux kernel 0-day build testing service~\cite{intel0day}, a continuous integration service developed by Intel.
This service performs both performance tests and build tests (including running various static analysis tools), and mails reports to patch developers on any detected regressions.
Accordingly, we only have access to information about the configurations in which regressions were detected, but these are also the configurations that have been the most useful.
We downloaded all of the available build-test messages from October 1, 2019 through August 27, 2022, resulting in 36,115 reports from the 0-day service containing configuration files.  Of these, the largest proportion are created using \texttt{make randconfig}, amounting to 22,702 configurations, or 63\% of the total.  This is followed by \texttt{make allyesconfig} at 5,551 (15\%), \texttt{make defconfig} at 2,708 (7\%), and \texttt{make allmodconfig} (analogous to \texttt{allyesconfig}, but trying to select as many modules as possible) at 2,446 (7\%).  The remaining 8\% were miscellaneous configuration files.
These results indicate that while the 0-day service does find the statement-coverage targeting configuration \texttt{allyesconfig} to be useful, most of its results are derived from \texttt{make randconfig}, which typically results in much smaller configurations.  But, as we show (Section~\ref{sec:evaluation}), \texttt{randconfig} provides little guarantee of patch coverage, even when run many times.

To help understand the challenges of performing continuous testing of highly-configurable software, we first overview the space of approaches to generating configuration files for testing such software.  Then we describe the Linux kernel build system, particularly focusing on the Kconfig language, and present a patch that raises configuration challenges.  
We finally consider how the Linux kernel build-system design impacts the problem of achieving build coverage of the lines affected by a given patch.

\subsection{Configuration Testing Approaches}

Fundamentally, the first challenge of testing of changes to highly-configurable software is to ensure that changes are not excluded from the tested binary, i.e., that the patched lines of code are compiled by the build system into the binary.
Performing continuous testing requires finding configurations that cover patches fast enough so that the testing infrastructure can keep up with the rapid pace of changes, which in Linux kernel development means hundreds of patches a day.
There is a trade-off in build time, which can require hours of processor time, and patch coverage.
Table~\ref{fig:comparison} compares state-of-the-art configuration file generation techniques along these two axes.

\newcommand{\STAB}[1]{\begin{tabular}{@{}c@{}}#1\end{tabular}}
\begin{table}
\caption{Comparing configuration testing approaches for use in continuous testing.}
\label{fig:comparison}
  \centerline{\small
    \def\arraystretch{1.27}
    \begin{tabular}{c@{\,\,\,}c|c|c|}
    \multicolumn{2}{c}{} & \multicolumn{2}{c}{\textbf{Patch Coverage}} \\
    
     \multicolumn{2}{c}{} & \multicolumn{1}{c}{Lower} & \multicolumn{1}{c}{Higher} \\\cline{3-4}
     
      \multirow{4}{*}{\STAB{\rotatebox[origin=c]{90}{\textbf{Build Time}}}}&
     \multirow{2}{*}{\STAB{\rotatebox[origin=c]{90}{Slower}}}&
     t-wise~\cite{taitest2002,10sampling} & \texttt{allyesconfig}~\cite{mainlinelinux} \\
     
     &&Combinatorial testing~\cite{cohen96} & VAMPYR~\cite{tartler2014} \\\cline{3-4}
     &
     \multirow{2}{*}{\STAB{\rotatebox[origin=c]{90}{Faster}}}&
     \texttt{randconfig}~\cite{mainlinelinux} & \cellcolor{light-gray} \\
     &&  \texttt{defconfig}~\cite{mainlinelinux} & \cellcolor{light-gray} \multirow{-2}{*}{\textbf{\toolname{}}} \\\cline{3-4}
     
  \end{tabular}}
\end{table}

Configuration file generation approaches that cover many feature interactions require generating many configuration files.  For instance, $t$-wise sampling ensures that each combination of $t$ configuration options is covered by some configuration file.  For 2-wise sampling, each of the four possible combinations of two options set to on or off needs to be covered by some configuration file for all pairs of configuration options.
As the Linux kernel has over 15,000 configuration options, 2-wise coverage would require considering a very large number of pairs, implying that even the most efficient algorithms cannot cover all interactions for the Linux kernel~\cite{10sampling}.
Such approaches are not designed for the problem of efficient patch coverage, but rather to test feature interactions, so repurposing them for patch coverage means very resource-intensive build times, due to the many configuration files needed.

Random configuration-file generation for testing is popular in industrial testing tools~\cite{intel0day,fixstats}, because each configuration file is much faster than using a statement-covering configuration file.
But our evaluation shows that individual randomly-generated configuration files have a low chance of covering patches.
Industrial tools compensate for a lack of coverage by generating multiple random configuration files, dozens in some cases~\cite{intel0dayslides,intel0day}.
But as our evaluation (Section~\ref{sec:evaluation}) also shows, adding more random configuration files has diminishing returns for patch coverage.  The tenth random configuration file adds less than a percent of additional coverage, and ten configuration file collectively only achieve 74.6\% patch coverage on average.
Adding more configuration files also inflates build time, because the total build time is proportional to the number of configuration files used.

Statement-maximizing approaches, such as \texttt{allyesconfig}, do have high patch coverage, since they attempt to cover as much of the source code as possible in one or a few configuration files.
But this good patch coverage, 88.5\% on average, comes at the cost of much slower build times, around four hours on average on a typical development workstation, compared to the default configuration or to \toolname{}'s configuration files, which take only around 20 minutes on average.
VAMPYR is a state-of-the-art statement coverage approach that improves on
\texttt{allyesconfig}~\cite{tartler2014}. Based on presence conditions for all of the statements
in a targeted code base, it employs a SAT solver to find a set of
configuration settings to cover the lines not covered by
\texttt{allyesconfig}, and then exploits the Linux kernel's \texttt{make
olddefconfig} to extend each resulting set of configuration settings with
default values to form a complete configuration.  VAMPYR thus produces a
set of configuration files covering more than \texttt{allyesconfig},
but at the cost of even slower build times, since it requires building
at least for \texttt{allyesconfig} as well as for its generated configuration files, and
the set of generated configuration files is not limited to what is needed
for a specific patch.
While evaluated for maximal statement coverage, VAMPYR's constraint covering approach can also be applied to only cover a patch's line constraints~\cite{tartler11b}.
\toolname{} builds on this constraint coverage approach by adding configuration repair, which automatically reconciles any existing configuration file, even very small ones, with patch coverage constraints, resolving the trade-off in build time and patch coverage by simultaneously enforcing patch coverage constraints and preserving most of the original configuration file's settings.

\subsection{Linux Kernel Configuration}

\begin{figure}
  \vspace{.2em}
  \hspace{.6em}
  \includegraphics[scale=.375]{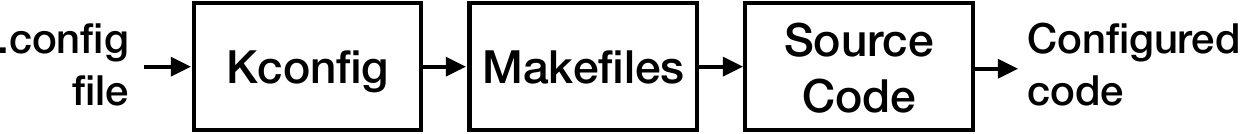}
  \vspace{.2em}
\caption{Build system components handling configuration.}
\label{fig:buildsystem}
\end{figure}

To help explain why the problem of finding an efficient, patch-covering configuration file is so difficult, let us first look at how the build system defines and uses configuration options.
The Linux kernel build system takes a configuration file as input and determines what files and lines of source code to compile into the kernel binary.
Figure~\ref{fig:buildsystem} shows the relevant components of the build system.

The first component is a collection of Kconfig files spread across the Linux kernel code base that formally describe the constraints on the configuration options relevant to each subsystem.
Kconfig is used to validate the input configuration file.
Figure~\ref{fig:example} shows a patch\footnote{\url{https://git.kernel.org/pub/scm/linux/kernel/git/torvalds/linux.git/commit/?id=8594c3b85171b6f68e34e07b533ec2f1bf7fb065}\label{fn:patch}} (Figure~\ref{fig:patch}) and Kconfig specifications (Figure~\ref{fig:kconfig}) for the options controlling the patched code.
The option with the simplest constraints is \texttt{PM} (Figure~\ref{fig:kconfig}, lines 11-12), which implicitly determines the value of the configuration variable \texttt{CONFIG\_PM}.
\texttt{PM} is declared as a Boolean (yes or no).  The associated prompt "Device power ..." indicates that the user will be asked with this prompt for the desired value.
\texttt{ARCH\_EXYNOS4} (line 16) is declared similarly, but it has a default value of \texttt{y} (yes, line 18).
In contrast, the constraints on \texttt{ARM\_GIC}, \texttt{ARM\_GIC\_PM}, and \texttt{GIC\_NON_BANKED}, indicate that, while these options are also Booleans, they cannot be specified directly by the user, as no prompt is provided.
Such configuration options may be declared to {\em depend on} the selection of another configuration option or can be {\em selected} by some other option.
\texttt{ARM\_GIC\_PM} depends on \texttt{PM} (line 5), and if it is selected, then it also selects \texttt{ARM\_GIC} (line 6).
Likewise, selecting \texttt{ARCH\_EXYNOS4} selects \texttt{GIC\_NON_BANKED}.
Finally, further constraints can be expressed using conditionals (e.g., \texttt{if ... endif}), as illustrated on lines 15-20.
A provided configuration file is checked to respect the various constraints specified by the Kconfig files, and is enhanced with any selected or dependent configuration options based on the options selected in the configuration file.
The result is a configuration that controls the rest of the build process.

The second component is the collection of Kbuild Makefiles spread across the Linux kernel code base that describe how to build and link the various subsystems.
As illustrated in Figure~\ref{fig:kbuild}, these Makefiles use the configuration variables to determine what files to include in the generated kernel.
For example, \texttt{irq-gic.c} is only compiled and included if \texttt{CONFIG\_ARM\_GIC} is set.

Finally, the third component is the source code itself.
Illustrated by line 4 of the patch (Figure~\ref{fig:patch}), source code may refer to configuration variables directly via \verb+#ifdef+s.
These \verb+#ifdef+s select the specific lines of code that will be included in the compiled kernel.
Changing the configuration file changes requires rebuilding the whole kernel, i.e., \verb'make clean', because make has no visibility over the \verb+#ifdefs+ used within C files.

\lstset{escapeinside={<@}{@>}}
\definecolor{aquamarine}{rgb}{0.5, 1.0, 0.83}
\definecolor{apricot}{rgb}{0.88, 0.71, 0.59}
\definecolor{aqua}{rgb}{0.0, 0.8, 0.8}
\definecolor{atomictangerine}{rgb}{1.0, 0.6, 0.4}

\definecolor{pred}{rgb}{0.5, 0.0, 0.0}
\definecolor{paqua}{rgb}{0.0, 0.4, 0.0}
\newcommand{\premoved}[1]{\texttt{\textcolor{pred}{#1}}}
\newcommand{\padded}[1]{\texttt{\textcolor{paqua}{#1}}}

\begin{figure}
\begin{subfigure}{\columnwidth}
\begin{lstlisting}[keywords=] % [morekeywords={if, bool, menuconfig, default, config, tristate, depends, on, select, help, source, endif}]
// from drivers/irqchip/Kconfig
config ARM_GIC
	bool
config ARM_GIC_PM
	bool
	depends on PM
	select ARM_GIC
config GIC_NON_BANKED
	bool

// from kernel/power/Kconfig
config PM
	bool "Device power management core functionality"

// from arch/arm/mach-exynos/Kconfig
if ARCH_EXYNOS
config ARCH_EXYNOS4
	bool "Samsung Exynos4"
	default y
	select GIC_NON_BANKED
endif
\end{lstlisting}
\caption{Kconfig specifications for options controlling the patched code, showing a few of the many dependencies.}
\label{fig:kconfig}
\end{subfigure}

\vspace{1em}

\begin{subfigure}{\columnwidth}
\begin{lstlisting}
// from drivers/irqchip/Makefile
obj-$(CONFIG_ARM_GIC)			+= irq-gic.o
\end{lstlisting}
\caption{Relevant build specifications for the patched file.}
\label{fig:kbuild}
\end{subfigure}

\vspace{1em}

\begin{subfigure}{\columnwidth}
\begin{lstlisting}[keywords=] 
--- a/drivers/irqchip/irq-gic.c
+++ b/drivers/irqchip/irq-gic.c
@@ -127,35 +124,27
 #ifdef CONFIG_GIC_NON_BANKED
<@\premoved{
-static void *gic_get_common_base(union gic_base *base)
}@>
<@\padded{
+static void enable_frankengic(void)
}@>
 {
<@\premoved{
-	return base->common_base;
}@>
<@\padded{
+	static_branch_enable(\&frankengic_key);
}@>
 }
 #else
<@\premoved{
-\#define gic_set_base_accessor(d, f)
}@>
<@\padded{
+\#define enable_frankengic()	do { } while(0)
}@>
 #endif
@@ -1165,7 +1149,7
<@\premoved{
-		gic_set_base_accessor(gic, gic_get_percpu_base);
}@>
<@\padded{
+		enable_frankengic();
}@>
\end{lstlisting}
\caption{Hunks from the patch to Linux source, edited for brevity.}
\label{fig:patch}
\end{subfigure}

\caption{An example patch to the Linux source and the configuration specifications controlling its buildability.}
\label{fig:example}
\end{figure}

\subsection{Motivating Example}

We next look at the same configuration constraints from the point of view of covering the changed lines of a patch.
The patch in Figure~\ref{fig:patch} affects the file \texttt{drivers/\-irqchip/irq-gic.c}.
It is formatted in the standard unified diff format~\cite{diffutils}, in which the \verb'-' prefix (lines~5, 8, 12, and 16) indicates a line to remove and the \verb'+' prefix (lines~6, 9, 13, and 17) indicates a line to add.
To cover the patch, a configuration must cause the modified file to be included in the build and ensure that all the changed lines are included in the build.

We start with the file.
Checking the Makefile in the same directory (Figure~\ref{fig:kbuild}) shows that building the file requires selecting the \texttt{ARM\_GIC} configuration option.
As previously noted, the user cannot select this option directly; instead, it is necessary to trace across multiple Kconfig files to discover that this option can be selected by the option \texttt{ARM\_GIC\_PM}. The latter option also cannot be selected directly but depends on \texttt{PM}.

We next turn to the lines changed within the file.
Indeed, simply ensuring that the build includes the file does not ensure that the build includes the changed lines, because some of these lines are under an \verb+#ifdef+.
The \verb+#ifdef+ involves the configuration option \texttt{GIC\_NON\_BANKED}; another search is needed to identify a selectable option that will cause this option to be selected.
But for this patch it is not sufficient to select \texttt{GIC\_NON\_BANKED}, because the patch modifies code under the \verb+#else+ as well.
The changes are thus mutually-exclusive and therefore covering all the changed lines requires at least two configuration files, one that selects \verb'GIC_NON_BANKED', and another that does not.

Still, even with all of the above collected information, the task of creating usable covering configuration files is not complete.
There are thousands of other options that need to be assigned, some of which may even influence whether the two identified options themselves are selectable.
Test cases that involve specific kernel features may introduce more configuration conflicts.

\textit{Assessment.}
The challenges in finding one or more configurations that cover a patch, as illustrated by the motivating example, come from the design of the build system.
Indeed, the build system is designed to take a configuration file and determine what lines of code to build, but not the other way around.
We can think of the build system as defining logical constraints on each line of source code~\cite{oh21}, and a configuration file as one solution to those constraints.
Determining the inverse, i.e., what configuration files build a certain line of code, is equivalent to a satisfiability problem, which is computationally expensive in the general case.
Finding what repairs to make to an existing configuration file requires determining what options directly control the patched lines, then reconciling those options with their dependencies and the settings in the existing file as much as possible, favoring patch coverage when settings in the existing file contradict the patch coverage constraints.

\section{The \toolname{} Algorithm}
\label{sec:algorithms}

\toolname{} automatically repairs an existing configuration file to ensure complete coverage of buildable code in a given patch.
It works in two steps: (1) discover a covering set of constraints for the lines changed (removed or added) by a patch and (2) find a set of changes to the existing configuration file that will satisfy these constraints.

\begin{figure}
  \centering
\includegraphics[scale=.85]{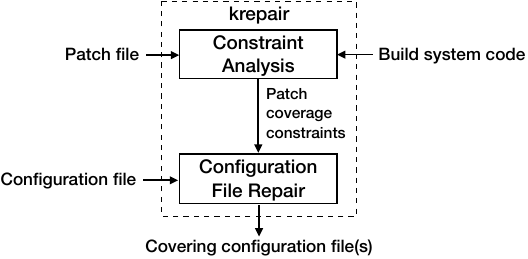}
\caption{Workflow of \toolname{}.}
\label{fig:process}
\end{figure}

Figure~\ref{fig:process} shows the high-level \toolname{} workflow.  Constraint Analysis takes as input the patch itself and the source code of the build system.
The output of the Constraint Analysis is a set of patch coverage constraints found by statically analyzing the build system code.
Configuration-File Repair then takes as input the existing configuration file to repair and produces one or more configuration files that are close to the input file but modified to cover the patch.

\begin{algorithm}[t]
\caption{$\textsc{Krepair}(\texttt{patchlines}, \texttt{configuration}, \texttt{commitid})$ - Repair an existing configuration file to cover the patch.}
\label{alg:repairall}
\begin{algorithmic}[1]
  \Require A list of (file, line) pairs in $\texttt{patchlines}$ from the patch and an existing configuration file $\texttt{configuration}$.
  \Ensure A set of repaired configuration files that cover that patch.
  \Function{\textsc{Krepair}}{\texttt{patchlines}, \texttt{configuration}, \texttt{commitid}}
  \State $\texttt{allrepaired} \gets \emptyset$

  \DoWhile
  \State $\texttt{current} \gets \textbf{true}$ \label{ln:startempty}
  \State $\texttt{covered} \gets \emptyset$
  \For {$\texttt{file,line} \in \texttt{patchlines}$}
     \State $\texttt{constraints} \gets \textsc{GetConstraint}(\texttt{file}, \texttt{line}, \texttt{commitid})$ \label{ln:getconstraints}
     \If {$\textsc{isSAT}(\texttt{current} \land \, \texttt{constraints})$}
        \State $\texttt{current} \gets \texttt{current} \land \texttt{constraints}$
        \State $\texttt{covered} \gets \texttt{covered} \cup \{(\texttt{file,line})\}$
    \EndIf
  \EndFor
  \If {$\texttt{covered} \neq \emptyset$}
    \State $\texttt{repaired} \gets \textsc{Repair}(\texttt{configuration}, \texttt{current})$
    \State $\texttt{allrepaired} \gets \texttt{allrepaired} \cup \{\texttt{repaired}\}$
    \State $\texttt{patchlines} \gets \texttt{patchlines} - \texttt{covered}$
  \EndIf
  \EndDoWhile{$\texttt{patchlines} \neq \emptyset \land \texttt{covered} \neq \emptyset$} \label{ln:endwhile}
  \State \Return $\texttt{allrepaired}$ \label{ln:return}
  \EndFunction
  \Function{\textsc{Repair}}{\texttt{configuration}, \texttt{constraint}}
  \State $\texttt{repair} \gets \texttt{configuration}$ \label{ln:repaironestart}
  \Repeat \label{ln:repaironerepeat}
  \State $\texttt{unsat} \gets \textsc{UnsatCore}(\texttt{repair} \land \texttt{constraint})$ \label{ln:repaironeunsatcore}
  \State $\texttt{repair} \gets \texttt{repair} - \texttt{unsat} $ \label{ln:repaironeremove}
  \Until {$\texttt{unsat} = \emptyset$} \label{ln:repaironeuntil}
  \State \Return $\textsc{SATSolve}(\texttt{repair} \land \texttt{constraint})$ \label{ln:repaironereturn}
  \EndFunction
\end{algorithmic}
\end{algorithm}

Algorithm~\ref{alg:repairall} describes \toolname{} in pseudo-code.
The algorithm takes as input a list of pairs of the file name and line number of those lines that are changed (added or removed) by a patch file.
The line number of an added line reflects its position after applying the patch.
The line number of a sequence of consecutive removed lines is the number of the line just preceding the removal after applying the patch.
The second input is the configuration file that needs repair.
The third input is the version of the code, as a commitid, that has had the patch applied to it.
The output is a \emph{set} of configuration files, since some patches may touch lines depending on mutually-exclusive configuration values.
For example, the patch in Figure~\ref{fig:patch} changes both arms of an \verb'#ifdef'.
Therefore, our algorithm cannot just conjoin all patch line constraints.
Instead, it tries to find a small set of satisfiable configurations that, together, cover the entire patch.
In practice, we find a single configuration for 97\% of patches, five or fewer for more than 99\% of patches, and 23 in the worst case.

\subsection{Constraint Analysis}

The algorithm first performs constraint analysis to partition the set of constraints controlling each patched line into subsets of constraints that do not contradict each other.
For this, it iterates repeatedly over the set of patched lines~(lines~3--18).
Each iteration starts with an empty constraint (line~\ref{ln:startempty}).
\toolname{} then iterates over each (file, line) pair~(lines~6--12), and greedily tries to cover as many pairs as possible within a single constraint (\texttt{current}).
This part of the algorithm draws from VAMPYR~\cite{tartler2014}, which achieves statement coverage by finding covering constraints.
Each (file, line) pair's constraint is provided by third-party tools that analyze the build system (\textsc{GetConstraint} on line~7).

If the (file, line) pair's constraint does not contradict the constraint accumulated so far (line~8), the algorithm updates the current constraint with that of the (file, line) pair (line~9).
It keeps track of which (file, line) pairs have already been accumulated (line~10), so that the algorithm will remove them from the set of candidates (line~16).
Once as many (file, line) pairs as possible have been accumulated, the algorithm repairs the configuration file according to the accumulated current constraint (line~14) and adds the result to a collection of repaired configurations (line 15).
\toolname{} stops trying to cover patch lines when either there are no more patch lines left to cover, or when it finds that no other patch lines can be covered by any configuration (line~\ref{ln:endwhile}).
The latter happens when (file, line) is unconfigurable, e.g., if it is configurable in another architecture or dependent on dead configuration options.

\textit{Optimizations.}
This algorithm relies on third-party constraint collection tools (line~7) and satisfiability solving to partition the set of patch line constraints (line~8), both of which are computationally expensive.
We make three optimizations.
The first optimization checks whether the patched line is inside any \#ifdef block.  If not, then there is no need to collect constraints from the source file; the line is always included if the file is included.
The second optimization checks whether a changed line is within the same set of \#ifdef blocks as an already-seen changed line.  In this case, there is no need to collect constraints for the current changed line.
These optimizations reduce the number of calls to the constraint-finding tool.
The third optimization targets nested \#ifdef blocks.
In this case, if the constraints for the inner block are satisfiable, then the constraints for the outer block must also be satisfiable.
Conversely, if the constraints for the outer block are not satisfiable, then the constraints for the inner block are also not satisfiable.
This optimization reduces the number of calls to the satisfiability checker.

\subsection{Configuration-File Repair}
\label{sec:cfr}

The repair part distinguishes \toolname{} from previous coverage approaches by automatically tailoring an existing configuration file so that it is patch covering without much change to the configuration.
The \textsc{Repair} function in Algorithm~\ref{alg:repairall} repairs the configuration file according to the given patch coverage constraints (lines~21--28).
It takes a configuration file and a constraint from \toolname{}'s constraint analysis and returns a configuration file close to the input file, but modified to satisfy the patch coverage constraints.
The repair algorithm repeatedly checks the configuration file against the patch coverage constraints and gradually removes configuration option settings until the configuration file satisfies the constraints.
Then, it repopulates any removed configuration option settings by taking a satisfying solution to the constraints.

The key trade-off in the repair algorithm is the computational complexity of finding the right settings to remove to satisfy the constraints of the patch while limiting the number of removals to keep the configuration file similar to the original.
A naive optimal algorithm for finding the minimal number of removals would be to check all combinations of setting removals against the constraints.
But this is prohibitively expensive, having an exponential number of satisfiability checks, i.e., the power set of thousands of configuration options.
A faster approach would be to remove some arbitrary number of options, check satisfiability after removing them, and repeat as needed.
But this approach might unnecessarily remove options that do not conflict with the patch coverage constraints.

Our algorithm has the better performance of the faster approach, while homing in on options that are preventing satisfiability more quickly.
For this, it repurposes feedback from the automated theorem prover, called an {\em unsatisfiable core}, to guide what settings to remove.
The unsatisfiable core is a (not necessarily minimal) subset of the original clauses that is still unsatisfiable~\cite{lynce04}.
By only removing settings in the unsatisfiable core, \textsc{Repair} gradually finds a subset of the configuration file options preventing satisfiability (lines~\ref{ln:repaironerepeat}-\ref{ln:repaironeuntil}).
Each new satisfiability check produces a new unsatisfiable core (line~\ref{ln:repaironeunsatcore}), which provides new removal candidates (line~\ref{ln:repaironeremove}).
Since \textsc{Krepair} only passes satisfiable patch coverage constraints to \textsc{Repair}, the unsatisfiable core always contains at least one configuration option as long as $\texttt{configuration} \land \texttt{constraint}$ is unsatisfiable, guaranteeing termination.
Finally, the missing constraints are repopulated by finding a satisfying solution to the reduced configuration file under the patch coverage constraints (line~\ref{ln:repaironereturn}).

\section{Implementation}
\label{sec:implementation}

\toolname{} is implemented as a command-line utility in
$\sim$3000 lines of python code.  It relies on third-party constraint-finding tools~\cite{fse17,gazzillo-pldi12,oh21} for \textsc{GetConstraint}.
\toolname{} runs from the root of a Linux kernel source tree, so it can identify the build system source files from its working directory.
It takes a patchfile and an existing configuration file on the command-line and produces output configuration files in the format expected by the build system.

\subsection{Processing Patch Files}

Linux kernel patches are represented in the unified diff format~\cite{diffutils}.
\toolname{} parses a patch using \texttt{whatthepatch}~\cite{whatthepatch} and converts the patch into a set of after-patch (file, line) pairs.
\toolname{} is line-oriented, so a patch that adds a new file requires checking coverage of all lines in the file.
\toolname{} does nothing when a file is simply renamed, as no lines are changed.  It also does nothing for removed files, as they are no longer buildable after the patch and therefore have no build constraints.  Removed lines, however, are considered changes just like added lines, since we can identify the configuration constraints affected by both by gathering the constraints for the file and any \texttt{\#ifdef} that contain them.

The build system only explicitly defines constraints for compilation units.
Therefore, \toolname{} provides limited support for patches to C header files, since headers are only included indirectly by other source files.
We use a simple heuristic to find covering constraints for lines in header files: \toolname{} assumes the header file has the same build constraints as the compilation units modified by the patch.
While this heuristic has some success in our evaluation, it means that patches that only modify header files are not supported, which we only encountered in 2\% of patches in our evaluation.

\subsection{Collecting Build Constraints}

\toolname{} uses third-party static analysis tools to collect build system constraints from each of the three build components: \texttt{kclause}~\cite{oh21} for Kconfig configuration specification constraints, \texttt{kmax}~\cite{fse17} for Kbuild Makefile constraints, and \texttt{SuperC}~\cite{gazzillo-pldi12} for preprocessor-level constraints in C source code.
Both \texttt{kclause} and \texttt{kmax} represent constraints in the SMT-LIBv2 format~\cite{smtlib2}, a standard representation of logical formulas for automated theorem provers.
\texttt{SuperC}, however, was not originally designed for reporting C preprocessor constraints, although its preprocessor collects them internally.
We forked the \texttt{SuperC} source code and added support for exporting the configuration constraints of all \texttt{\#ifdef} ranges and their constraints from a given source file in the SMT-LIBv2 format.

\toolname{}'s constraint collection module interfaces with all three tools, providing python wrappers around each to implement the \textsc{GetConstraints} function from Algorithm~\ref{alg:repairall}.
For a given (file, line) pair, \textsc{GetConstraint} collects the results from each of the three tools, and then conjoins them into a single constraint for the line.

\subsection{Improving Performance}
\label{caching}

Our algorithm only needs access to per-line build constraints for a given patch.
But the tools we use to collect constraints were designed to run on the
entire build system source files. For instance, \texttt{kclause} takes as input the entire 140,000+ lines of Kconfig specification and converts it to about 60,000 logical clauses all at once.
The tools are thus time-consuming to run, with \texttt{kclause} typically taking 2-3 minutes, \texttt{kmax} 10-15 minutes, and \texttt{SuperC} about a minute or less on commodity hardware, all to get a single line's constraints, depending on the Linux kernel version and the target architecture.
To reduce the cost of constraint collection for a single patch, we modified the \texttt{kmax} interface to support collecting per-file constraints on-demand.
We also modified \texttt{SuperC} to emit per-line constraints for the entire source file.

The Kconfig configuration specification is a single large constraint for each of the supported architectures.
Since the Kconfig specifications do not change with every patch, \toolname{} manages an on-disk cache of Kconfig constraints indexed by a unique identifier of the Kconfig version that will be reused as long as the Kconfig specification has not changed.
Similarly, Kbuild Makefiles, which define constraints on source files, only need to be collected once for a given file until the Makefile source code changes.

\subsection{Implementing Repair}

As with build constraints, the configuration file is represented as a set of SMT-LIBv2 constraints.
\toolname{} has functions to parse configuration files into constraints and to deparse satisfying solutions to constraints back into the configuration file format.
The implementation of the repair algorithm (Algorithm~\ref{alg:repairall}) uses the z3~\cite{z3} automated theorem prover to check satisfiability (\textsc{isSAT}), find an unsatisfiable core (\textsc{UnSATCore}), and get a satisfying solution (\textsc{SATSolver}).
z3 is not guaranteed to provide a minimal unsatisfiable core, but we find that the resulting cores are small enough that our repairs incur little change, less than 2.23\% change for 99\% of repairs.

\section{Empirical Evaluation}
\label{sec:evaluation}

We evaluate \toolname{} on a representative sample of Linux kernel patches, measuring how well it ensures that configuration files cover patches while keeping the build times fast enough for continuous testing.
We study patches from the Linux kernel, because it is large, highly-configurable, very actively developed, and used in critical computing infrastructure.

\subsection{Experimental Setup}

{\em Sampling patches.}
We have taken a random sample of 507 patches out of the approximately 71,000 patches from one recent whole year (2021/09/19--2022/09/18) of Linux kernel development, which provides a 5\% margin of error with a 98\% confidence level.
We performed sampling by cloning the mainline Linux kernel repository~\cite{mainlinelinux} and using \verb'git log' on the above date range.  We exclude merge commits, which typically do not change code, and include only patches to buildable kernel source files, which excludes documentation text files, example programs, build tools, and header files.
Such files are not covered by any configuration file, since they do not get compiled and linked into the kernel binary, although header files may be indirectly covered when they are included in other kernel source files.

{\em Configuration file collection.}
Our baseline for a fast-building configuration file is the \emph{default configuration file} distributed with the kernel source for the x86 platform.
This configuration file, created with \texttt{make defconfig}, is a small, quick-to-build kernel configuration frequently recommended as a starting point for building the kernel~\cite{yaghmour03,syzkaller} and frequently used in testing~\cite{intel0day,syzkaller}.
Compiling with \texttt{defconfig} is fast, because it enables relatively few options, therefore covering very little of the code.
Our baseline for a statement-covering configuration file is \texttt{make allyesconfig}, which attempts to enable as many configuration options as possible.
Although mutual exclusion among configuration options prevents coverage of all code, it still covers the large majority of code (and therefore patches), at the cost of much longer build times.
While VAMPYR improves on \texttt{allyesconfig}'s coverage, it always builds \texttt{allyesconfig} plus additional configuration files~\cite{tartler2014}.  Thus, it always causes even higher build times than \texttt{allyesconfig}.
We repair \texttt{defconfig} by applying \toolname{} to the configuration file to ensure patch coverage, which we expect to achieve the best of both worlds, the high patch coverage of \texttt{allyesconfig} and the much faster build times of \texttt{defconfig}.
Additionally, we evaluate the patch coverage capability of randomly generated configuration files, which is a lightweight method to attempt to increase code coverage.
When evaluating random configuration testing, we use Linux's built-in random configuration file generator (\texttt{make randconfig}).
For evaluating how much change in the configuration file \toolname{} causes, we compare against \texttt{defconfig} as well as \texttt{allnoconfig}.  \texttt{allnoconfig} is the Linux kernel's minimal configuration file that disables most configuration options, and we use it as an extreme test case for \toolname{} since it covers few patches.

{\em Metrics collection.}
To collect metrics for a configuration file on one patch from the sample, we first check out the kernel using the patch's commit ID.
We configure and build the kernel using each of the tested configuration files, collecting patch coverage and build time.
Patch coverage is evaluated by saving the source code of the patched files after they are configured by the build system, i.e., the preprocessed \emph{.i} files, and checking which lines of the patch have been included or excluded by the build system.
We quantify patch coverage as the ratio of changed (added or removed) lines included in the build over the total number of changed lines in the patch.\footnote{Non-source files, such as documentation or example source code, are not considered in the total lines, since they are never compiled into the binary.}
When the patch adds an entirely new file, we consider each line in the file as added by the patch.
When the patch removes a file, we consider it as having no lines, since there is no way to build the affected lines of code.
Removed lines from an existing file, however, are measured by looking at whether the enclosing \texttt{\#ifdef} block around the removed line (or the entire file, if there is no \texttt{\#ifdef}) is included by the configuration file, since that controls the inclusion of the change and does get built.
We quantify build time by recording the wall clock time of the build process (\texttt{make}) using the UNIX \verb'time' utility.

{\em Parallel builds.}
\texttt{make} supports parallel builds with the \texttt{-j} flag, which allows {\tt make} to compile source files in parallel when there are no dependencies between them.
Parallel builds do not affect patch coverage or build size, since the same files are compiled with the same configuration file; they only affect build time.
For our build time comparisons, we use eight concurrent build threads, since this reflects the power of modern developer laptops, as well as a single thread to record sequential build time.
We show the effect on performance of parallel builds in Section~\ref{sec:rq1} by comparing the sequential and parallel build times of the kernels in the sample.

{\em Cross-compilation.}
We run our experiments on a 64-bit x86 machine, but some patched source code can only be built for non-x86 architectures.
The patch format does not force the developer to specify for what architecture the code is meant to be built; indeed, one patch may touch code built for multiple architectures.
\toolname{}, however, can detect for which architecture(s) a patch is built by exploring the space of configuration constraints for each architecture's Kconfig specification.
To build code for other architectures, we perform cross-compilation using the \texttt{make.cross}\footnote{https://github.com/fengguang/lkp-tests/blob/master/sbin/make.cross} utility, which automates downloading and installing build tools for most other architectures (2 patches are from an architecture that is not supported, so we could not automatically evaluate their patch coverage).
We find that 11\% of patches in our sample require cross-compilation.

{\em Computing platform.}
All experiments were run on a server with dual AMD EPYC 7742 64-Core Processors and 512GB of RAM running Ubuntu 22.04.03 LTS.
Since this machine allows for high parallelism and our builds are only using either one or eight threads, we parallelized the experiment scripts.
All experiment scripts are available in the code repository~\cite{kmaxrepo} as well as the artifact archive~\cite{artifact} under \texttt{scripts/krepair_evaluation/paper/}.

\subsection{Research Questions}

We ask the following research questions (RQs) to evaluate \toolname{}:
\begin{enumerate}
\item[RQ1] (Efficient Patch Coverage) Does krepair produce configuration files with high patch coverage and fast build times?
\item[RQ2] (Performance) How fast is krepair?
\item[RQ3] (Configuration Preservation) How well does krepair preserve the settings of the repaired configuration file?
\item[RQ4] (Random Testing) How well does random configuration testing cover patches compared to \toolname{}?
\item[RQ5] (Build Errors) Can \toolname{} help reveal build errors?
\end{enumerate}

\subsection{RQ1: Efficient Patch Coverage}
\label{sec:rq1}

\begin{figure}
  \hspace{.6em}
  \includegraphics[scale=.6]{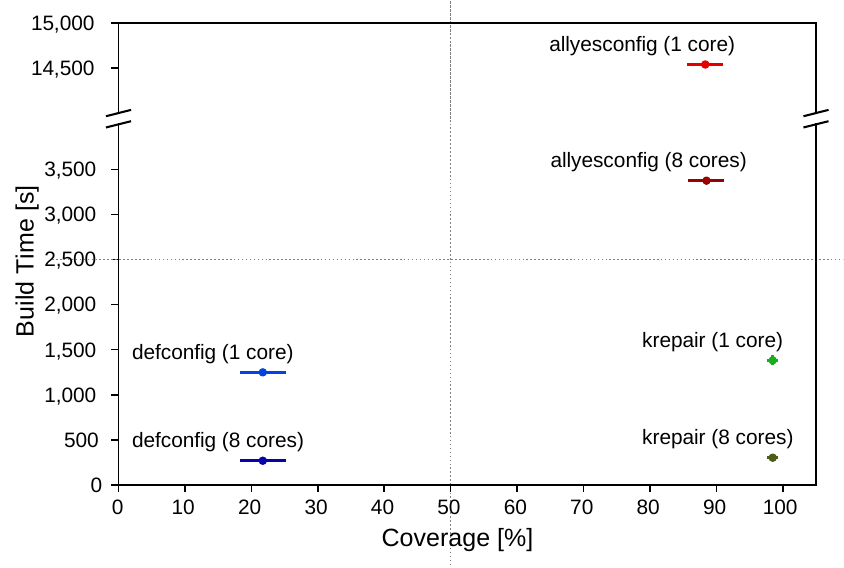}
\caption{Patch coverage plotted against build times.}
\label{fig:coverage}
\end{figure}

In our experiment, we collect patch coverage and build-time metrics when building each commit in the sample using configuration files made using \texttt{make defconfig}, \texttt{krepair} to repair \texttt{defconfig}, and \texttt{make allyesconfig} (\texttt{randconfig} will be evaluated in RQ4).
Figure~\ref{fig:coverage} compares patch coverage (x-axis, higher is better) to the build time (y-axis, lower is better).
Each point is the average of all patch coverage percentages and the average of all build times in seconds across the entire sample of patches, excluding failed builds of which there were 27.
The error bars are the 95\% confidence interval.
For each of the three configuration file generators, we plot both the single-core build time as well as the eight-core parallel build time, resulting in six total points.

Since \toolname{} is intended to preserve fast build times while still covering patched code, we consider it a success if it can simultaneously outperform \texttt{defconfig}'s patch coverage and \texttt{allyesconfig}'s build time.
The results show that repairing \texttt{defconfig} with \toolname{} results in much higher patch coverage, about 4.5x more, and even produces higher patch coverage than \texttt{allyesconfig}.
In contrast, the build times of the repaired \texttt{defconfig} are substantially faster, about 10.5x faster, remaining comparable to \texttt{defconfig} even after repair.
The narrow error bars show that these results from our sample are statistically significant.
In short, repairing configuration files with \toolname{} achieves nearly complete patch coverage, while adding little additional build time.  Parallel builds reduce build times for all configuration files roughly proportionately.

\subsection{RQ2: Performance}
\label{sec:performance}

\begin{figure}
  \centerline{
  \includegraphics[scale=.6]{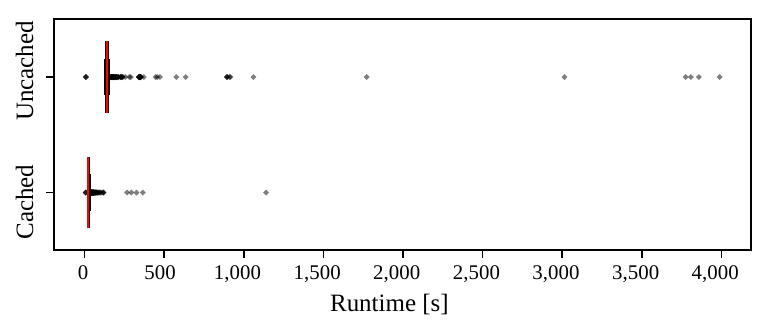}
  }
\caption{\toolname{}'s cached and uncached runtime.}
\label{fig:performance}
\end{figure}

We measure \toolname{}'s repair runtimes for \texttt{defconfig} with and without caching (Section \ref{caching}).
Figure~\ref{fig:performance} is the distribution of times for each patch in the evaluation sample.
The Uncached row measures timing when the build-constraints cache is cold and has no prior build constraints cached.
The Cached row assumes the build system constraints for each patch have already been cached.

The boxes for the interquartile ranges for both cached and uncached and the lines for the confidence intervals are so narrow, relative the maximum runtime, that they appear to be single a red line on the graph.
In other words, the large majority of \toolname{} runtimes take only a few minutes, even without caching, and there are only a small number of outliers that take up to an hour in rare cases.  Caching provides a substantial benefit, reducing the maximum runtime by 71.4\% to around 19 minutes and less than one minute for 93.8\% of runs.

\subsection{RQ3: Configuration Preservation}

\newcommand{\cc}[1]{\cellcolor{red!60!white!#1}}
\renewcommand{\cc}[1]{}

\toolname{} maintains the build time of the original configuration file because it keeps the set of changes to an existing configuration file small when ensuring patch coverage.
Therefore, by starting with a fast-building configuration file, such as the default configuration file, the build time is largely preserved to be fast, while patch coverage increases, as shown in Figure~\ref{fig:coverage}.
We measure how much change \toolname{} incurs to ensure patch coverage to demonstrate its effectiveness at preserving the original configuration file.
We define change as the number of configuration options that differ in their setting between two configuration files divided by the total number of configuration options available.

In addition to evaluating \toolname{}'s repair of \texttt{defconfig}, we also evaluate its effectiveness on \texttt{allnoconfig}.
\texttt{allnoconfig} is the Linux kernel's minimal configuration file that disables most configuration options and therefore is an extreme test case for \toolname{} since it covers few patches.

Table~\ref{tab:rq3} is the percentile distribution of change, as a percentage of all configuration options available in the kernel, incurred by \toolname{} when repairing both \texttt{defconfig} and \texttt{allnoconfig}.
For \texttt{defconfig}, it changes no more than 10\% of the configuration options at the extreme, and no more than 1.53\% for 99\% of the patches in the sample.

\begin{table}[]
\centering
\caption{
Distribution of percent change incurred by \toolname{}.
}
\begin{tabular}{l|rrrrr}
\label{tab:rq3}
\textbf{Comparisons} & \textbf{Min} & \textbf{25th} & \textbf{Median} & \textbf{99th} & \textbf{Max} \\
\hline
allnoconfig        & \cc{0}{0.46\%}  &  \cc{1}{1.06\%}  &    \cc{1}{1.48\%}    &  \cc{3}{2.23\%}    &  \cc{6}{4.98\%}   \\
defconfig        & \cc{0}{0.14}\%   & \cc{1}{0.21}\%  &    \cc{1}{0.27}\%    &  \cc{3}{1.53}\%    &  \cc{6}{9.52}\%   \\
\end{tabular}
\end{table}

Similarly, \toolname{} finds that only a relatively small amount of change is needed for \texttt{allnoconfig}, in spite of it having few options enabled, in order to cover the patches in the sample.
In the majority of cases, \texttt{allnoconfig} requires more changes than \texttt{defconfig}, which is to be expected, given how few options the configuration selects initially.
But at the extreme cases, \texttt{defconfig} requires more changes; when a configuration file has comparatively more options enabled, there is a chance that the enabled options contradict the dependencies needed for covering the patch, requiring \toolname{} to first disable these options, causing higher total change.

\subsection{RQ4: Random Testing}
\label{sec:randconfig}

\begin{figure}
  \hspace{.6em}
  \includegraphics[scale=.6]{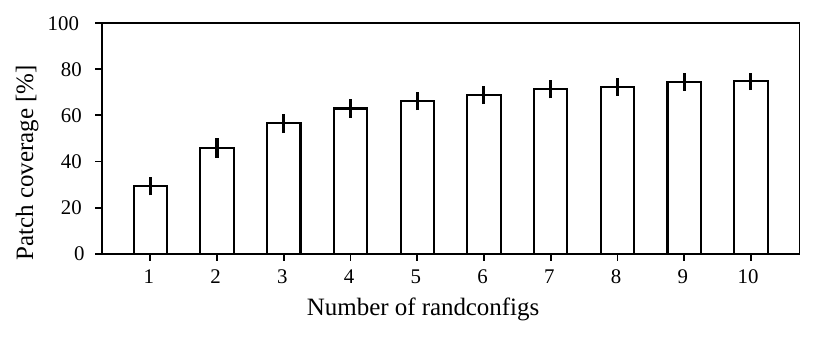}
\caption{Aggregate coverage of random configuration files.}
\label{fig:rq4}
\end{figure}

Random configuration file testing is commonly used in continuous configuration testing, so we also evaluate how the patch coverage from multiple random configuration files compares to \toolname{}'s coverage.
For each patch in our sample, we generate a series of ten random configuration files, measuring their aggregate patch coverage.
Figure~\ref{fig:rq4} shows the aggregate patch coverage of 1 to 10 randomly generated configuration files for each patch in the sample.
The error bars show the 95\% confidence interval of the average for the sample.

Generating a single random configuration file results in low patch coverage, at 29.2\%.
Increasing the number of configuration files increases coverage, but with diminishing returns; the amount of additional coverage plateaus at 9 random configurations, which in aggregate only cover around 74.4\% of patches on average.
In contrast, \toolname{} achieves much higher patch coverage, 98.5\% on average, and almost always with a single configuration file instead of having nine configuration files requiring more build time.

\subsection{RQ5: Build Errors}

\begin{table}
\caption{Build errors found when evaluating \toolname{}.}
\label{tab:builderrors}
\begin{tabular}{ll}
\textbf{Build Error} & \textbf{Commit ID(s)} \\
\hline
Warnings treated as errors & 3f977c57, c1318b39, 6ece49c5, c974f755, 5dee8bb8 \\
Linker error & 0258cb19, e0905322, 16dd1fbb, dfbdcda2, 661c399a \\
Implicit function declaration & f9135821, b5054161, ae9fd76f, 4a46e5d2 \\
Incompatible pointer type & 800fe5ec, c8992cff \\
Frame size error & 8763e4c1 \\
Undeclared variable & bce84458 \\
\end{tabular}
\end{table}

In our experiments, we found that some configuration files we generated for patch coverage failed to build.
This is unsurprising, since it is infeasible for developers to build all variants of the kernel.
While \texttt{defconfig} and \texttt{allyesconfig} are frequently tested and typically do not trigger build errors in released code (code given a version number), small variations in a configuration file can expose new bugs.

We found 18 build errors due to several bugs.
Table~\ref{tab:builderrors} lists the build errors found, with the commit ID(s) through which they were found.
These bugs were not introduced by the corresponding commits, but were present at the checkout of the commit.
Ten were due to missing symbol declarations (linker errors, implicit function declarations, undeclared variables).
Missing declarations occur in highly-configurable software when the declaration of a symbol is disabled by one configuration option and the use of the symbol is enabled by another.
Five build errors were due to \texttt{-Werror} being enabled by the configuration file, causing compiler warnings (which by default do not halt compilation) to trigger a compiler error.
Two commits failed due to pointer type mismatches, and one due to a display mode subsystem error: ``the frame size of 2112 bytes is larger than 2048''.
We patched one of two build errors still replicable in the recent v6.1-rc8 kernel.  This patch has been accepted for inclusion in mainline Linux, while we plan to patch the other.
The rest of the bugs were no longer in the most recent kernel.

Since we build non-x86 patches on x86 hardware, we could not cross-compile some of the patches due to limitations of our cross-compilation tooling, \texttt{make.cross}.
The \texttt{make.cross} script does not support the newly-added \texttt{loongarch} architecture and some cross-compilers had incompatibility, for instance, reporting unexpected assembly opcodes.  These cross-compilation problems prevented us from building seven patches: 8c4d1647, 0b452520, 7eafa6ee, 44c14509, 6982dba1, f62b7626, 54cfa910.
\toolname{} determined the parisc 32-bit architecture for two commits, 53d862fa and db2b0d76, while the configuration files instead required the parisc 64-bit cross-compiler, which is not available with \texttt{make.cross}.

\section{Threats to Validity}

\textit{Internal validity.}  Since \toolname{} relies on existing constraint collection tools~\cite{oh21,gazzillo-pldi12,fse17}, any limitations of these tools limit \toolname{}.
Specifically, these tools only collect constraints from the build system, while other sources of constraints are not supported, such as run-time uses of configuration options, i.e., with C conditionals instead of \#ifdef.
Additionally, header file inclusion constraints are also not available from these tools, though future work on constraint collection could yield analyses that discover all possible ways a header file is included across the entire kernel source.
Even without the above limitations, 100\% patch coverage may not necessarily be possible in all cases, as some patches change dead code in \verb+#ifdef 0+ blocks, which can never be included in any build.

\textit{External validity.}  We evaluate \toolname{} on only one software system, the Linux kernel source code, albeit one of the largest and most highly-configurable open-source software products.
While our implementation is targeted to the Linux build system, this build system is also used by numerous systems and embedded open-source projects (BusyBox, coreboot, zephyr, etc.).  \toolname{}'s algorithm, however, is independent of any particular build system, since it operates on any configuration constraints extracted from the software product.
\toolname{} focuses on the problem of patch coverage to enable more efficient continuous integration, since current approaches cannot even guarantee that patched lines are built.
Testing all the effects of a patch, however, goes beyond just line coverage; succeeding in compiling does not guarantee a test suite will execute the code without additional analysis.
Patch coverage is the first step to any kernel testing, so we are exploring future work on combining our repair approach with kernel fuzz-testing~\cite{syzkaller}, change impact analysis~\cite{ryder01}, configuration interaction testing~\cite{yuangui2011}, and other testing strategies~\cite{kernelci}.

\section{Related Work}

To the best of our knowledge \toolname{} is the first technique to repair Linux configuration files for patch coverage.
We highlight work related to \toolname{} and that addresses related problems in the domain of configurable software.

\textit{Configuration coverage.}
JMake~\cite{lawall17} is a previous attempt to find a configuration that covers a patch. However, it tries only a fixed set of standard configurations. JMake also introduces a mutation-based approach to determining if a line of code is covered.
Acher et al.~\cite{acher2019learning} explore the effect of configurations on compiled Linux kernel sizes, and compare machine learning approaches for predicting compiled size from configurations. They also explore small Linux builds and their use cases.
Tartler et al.~\cite{tartler11b} introduce a metric for how much of the source code is covered by a configuration.
Motivated by the results obtained for this metric, Tartler et al. \cite{tartler2014} created VAMPYR, a statement-maximizing approach discussed in Section~\ref{sec:overview}.
Note that VAMPYR is an older tool and no longer maintained. It only supports up to around Linux 3.2 (released in 2012).

\textit{Configuration constraint finding.}
\toolname{} takes inspiration from prior work on collecting constraints from Linux build-system code to get patch-covering constraints during repair.
Several prior works extract constraints from Kconfig specifications by translating Kconfig language constructs into logical formulas or feature models~\cite{kconfigreaderarxiv,oh21,dietrich12a,sincero2010efficient,she2013feature}.
Kbuild Makefile analysis collects logical constraints using both static and dynamic program analyses~\cite{nadi2012,fse17,berger2010feature}.
Several C preprocessor static code analyzers model configuration constraints in logic \cite{gazzillo-pldi12,kastner11,garrido2005analyzing,xref,spinellis2010cscout}, albeit for parsing, type-checking, refactoring, bug-finding, etc., rather than constraint extraction.
Prior work on localizing configuration constraints per-line aggregates constraints from multiple sources, including Kconfig, Kbuild, and the C preprocessor~\cite{gazzillo2018localizing,pclocator}, although it does not scale to the Linux build system.
Collecting line constraints is not enough to create a valid Linux kernel configuration file, due to the need to additionally incorporate basic system functionality and the possibility of conflicting constraints.
While \toolname{} is the first tool we know of to automatically achieve patch coverage, there are applications of configuration constraints in prior work to other software engineering problems, including attack surface reduction~\cite{kuo20,kurmus2013attack}, dead code elimination~\cite{tartler11a}, statistical analysis of build errors~\cite{acher2019learningb}, configuration tracing~\cite{franz2021configfix} and configuration specification bug-finding~\cite{oh21}.

\textit{Analyses for other configuration systems.}
The Puppet deployment configuration language has formal verification by Shambaugh et al. \cite{rehearsal}, automated repair by Weiss et al. \cite{weiss2017tortoise}, and a formal model of the system call trace by Sotiropoulos et al. \cite{sotiropoulos2020practical}.
Formal models are also used for system configuration script and resource usage by Hanappi et al. to test if a system is recoverable \cite{hanappi2016asserting}, as well as by Bouchet et al.~\cite{bouchet2020block} to check for public access to Amazon S3 instances. Horton et al. \cite{horton2019dockerizeme} infer dependencies from Python code snippets to produce Docker specifications. Sun et al.~\cite{sun20} introduce ctests to detect potential system failures from configuration changes. Cheng et al. do configuration test case prioritization~\cite{cheng21}.  Tamrawi et al. \cite{tamrawi-symake} introduce SYMake, which performs static analysis of Makefiles to detect errors like cyclic dependencies and can aid in refactoring. MAKAO, by Adams et al. \cite{adams07}, can be used to create graphs of Makefile dependencies for visualization.
Zhang and Ernst explore retaining system behavior after changes~\cite{zhang14}.

\textit{Random sampling for configuration testing.}
While random configuration testing is difficult to scale to the Linux build system~\cite{10sampling}, sophisticated testing approaches for smaller systems include genetic algorithms~\cite{guo11}, pair-wise feature selection~\cite{taitest2002}, and combinatorial interaction testing~\cite{cohen08,oh19,yilmazmoving2014}.

\textit{Tracking evolution of Linux patches.}
\toolname{}'s evaluation looks at a sample of patches over time.  Prior work has also measured how the configuration system evolves over time, in particular how they relate to code size~\cite{lotufo2010evolution}, what patterns are in the mapping between options and implementation~\cite{passos2016coevolution}, how configuration options change over time~\cite{dintzner2017analysing}, how patches affect configuration specifications~\cite{dintzner2018fever}, and how changes of Kconfig impact source code~\cite{ziegler2016analyzing}.

\textit{Fixing configuration errors.}
A related but distinct line of work addresses the problem of fixing configuration errors~\cite{white08,xiong12,wang13}, such as those that appear after code evolution.  In using the term ``repair'' in our work on \toolname{}, we are referring to automatically modifying a valid configuration file to remedy its lack of patch coverage.  But we do not address the problem of fixing erroneous configuration files.

\section{Conclusion}

We have shown how \toolname{} achieves much higher coverage of patches in kernel builds via automated repair of configuration files.
Its algorithm's design and implementation balance the expense of satisfiability with tool performance to achieve patch coverage comparable to maximal configuration files while preserving most configuration options settings from the repaired configuration file.
\toolname{} keeps build times fast while retaining patch coverage, potentially reducing the energy costs of configuration testing which relies heavily on building many randomly-generated configuration files.
Our evaluation shows that \toolname{} achieves 4.5x more patch coverage than default configuration files with 10.5x less build time than maximal configuration files on a statistically-significant sample of Linux kernel patches.
For future work, we plan to extend \toolname{} to other problems, such as fuzz-testing, change impact analysis, configuration bisection, and other testing and analyses for highly-configurable software.

\section{Data Availability}

The \toolname{} tool has been released as free-and-open-source software as part of the kmax tool suite~\cite{kmaxrepo} and has also been archived on Zenodo~\cite{artifact}.  The scripts to run experiments and the resulting data has been archived on Zenodo~\cite{dataset}.

\section*{Acknowledgments}
We would like to thank the anonymous referees for their
valuable comments.
This work is supported in part by the
\grantsponsor{NSF}{National Science Foundation}{}
under grant \grantnum 
{NSF}{CCF-1941816}.

\bibliographystyle{ACM-Reference-Format}
\bibliography{references}

\end{document}